\documentclass[american,english,aps,twocolumn,a4paper,10pt,superscriptaddress,nofootinbib,nobibnotes,altaffillsymbol,showpacs,showkeys]{revtex4}
\usepackage[T1]{fontenc}
\usepackage[utf8]{inputenc}
\usepackage{color}
\usepackage{babel}
\usepackage{amsmath}
\usepackage{amssymb}
\usepackage{graphicx}
\usepackage[unicode=true,
 bookmarks=true,bookmarksnumbered=false,bookmarksopen=true,bookmarksopenlevel=2,
 breaklinks=false,pdfborder={0 0 0},backref=false,colorlinks=true]
 {hyperref}
\hypersetup{pdftitle={Entanglement Efficiencies in PT-Symmetric Quantum Mechanics},
 pdfauthor={Christian Zielinski and Qing-hai Wang},
 pdfsubject={Entanglement Efficiencies in PT-Symmetric Quantum Mechanics},
 pdfkeywords={entanglement generation, entanglement efficiency, entanglement entropy, PT-symmetric quantum mechanics},
 linkcolor=BlueViolet,anchorcolor=BlueViolet,citecolor=BlueViolet,filecolor=BlueViolet,urlcolor=BlueViolet}

\makeatletter
\@ifundefined{textcolor}{}
{%
 \definecolor{BLACK}{gray}{0}
 \definecolor{WHITE}{gray}{1}
 \definecolor{RED}{rgb}{1,0,0}
 \definecolor{GREEN}{rgb}{0,1,0}
 \definecolor{BLUE}{rgb}{0,0,1}
 \definecolor{CYAN}{cmyk}{1,0,0,0}
 \definecolor{MAGENTA}{cmyk}{0,1,0,0}
 \definecolor{YELLOW}{cmyk}{0,0,1,0}
}

\usepackage[usenames,dvipsnames]{xcolor}
\usepackage{lmodern}
\usepackage{bbold}

\allowdisplaybreaks

\makeatother

\begin{document}

\title{Entanglement Efficiencies in $\mathcal{PT}$-Symmetric Quantum Mechanics}

\author{Christian Zielinski}

\email{zielinski@thphys.uni-heidelberg.de}

\selectlanguage{english}%

\affiliation{Institut für Theoretische Physik, Universität Heidelberg,\\
Philosophenweg 16, 69120 Heidelberg, Germany}

\author{Qing-hai Wang}

\email{phywq@nus.edu.sg}

\selectlanguage{english}%

\affiliation{Department of Physics, National University of Singapore,\\
117542, Singapore}

\date{\today}
\begin{abstract}
The degree of entanglement is determined for an arbitrary state of
a broad class of $\mathcal{PT}$-symmetric bipartite composite systems.
Subsequently we quantify the rate with which entangled states are
generated and show that this rate can be characterized by a small
set of parameters. These relations allow one in principle to improve
the ability of these systems to entangle states. It is also noticed
that many relations resemble corresponding ones in conventional quantum
mechanics.
\end{abstract}



\maketitle

\section{Introduction}

In conventional quantum mechanics, one demands that the Hamiltonian
$H$ generating the time evolution has a real spectrum and that the
corresponding time evolution operator $U$ is unitary. These conditions
are fulfilled if the Hamiltonian is Hermitian, i.e.~$H=H^{\dagger}$,
which is usually considered an axiom of quantum mechanics. However,
the condition of Hermiticity can be weakened. In the class of so-called
$\mathcal{PT}$-symmetric Hamiltonians, one can ensure real eigenvalues
and a unitary time evolution even for explicitly non-Hermitian Hamiltonians
\cite{PhysRevLett.80.5243,bender:2201}. A thorough review of the
foundations of $\mathcal{PT}$-symmetric quantum mechanics can be
found in \cite{0034-4885-70-6-R03}.

In the following we will investigate entanglement phenomena in bipartite
systems within this framework. An entangled state is a quantum state
where two or more degrees of freedom are intertwined, so that they
are not independent anymore. In this context a couple of historical
discussions took place and gave deep insights into the nature of quantum
mechanics, like the \textit{Einstein-Podolsky-Rosen} paradox \cite{PhysRev.47.777}.
These states have a wide range of applications, for example in quantum
information theory and quantum computing. The question of entanglement
generation and entanglement efficiencies for conventional quantum
mechanics was addressed earlier in \cite{PhysRevLett.87.137901}.

Using $\mathcal{PT}$-symmetric quantum mechanics to describe entanglement
phenomena was also done in \cite{springerlink:10.1007/s12043-009-0101-0}.
Especially for the case of bipartite systems, relations for the degree
of entanglement of given states and the entanglement capability of
certain systems could be found. Although we confirmed many of these
results, we have some different and new findings. For a particular
initial state, as well as for general states, we give relations between
the efficiency of the system to generate entangled states and the
parameters of the Hamiltonian describing the dynamics of the system.

We will first introduce a measure for entanglement and a certain class
of $\mathcal{PT}$-symmetric Hamiltonians. We then quantify the degree
of entanglement of an arbitrary and of a generalized \textit{Einstein-Podolsky-Rosen}
state. Subsequently, we are dealing with the question of entanglement
generation of a particular $\mathcal{PT}$-symmetric state and generalize
for arbitrary states. The question we try to answer is how to characterize
the rate of entanglement generation for this class of systems. \global\long\def\tr{\operatorname{tr}}
 \global\long\def\matrixOne{\mathbb{1}}
 \global\long\def\ii{\mathsf{i}}

\section{Bipartite systems}

\subsection{A measure for entanglement}

Let $\mathcal{H}_{i=1,\ldots,N}$ denote a set of Hilbert spaces.
We call a state of a composite system $\mathcal{H}=\mathcal{H}_{1}\otimes\dots\otimes\mathcal{H}_{N}$
entangled, if there is no decomposition of the form $|\Psi\rangle=|\chi_{1}\rangle\otimes\dots\otimes|\chi_{N}\rangle$
with suitable $|\chi_{i}\rangle\in\mathcal{H}_{i}$. In the following
we restrict ourselves to bipartite systems, i.e.~$N=2$.

A measure of entanglement are the entropies
\begin{equation}
E(\Psi)=-\tr_{1}\left(\rho_{1}\log_{2}\rho_{1}\right)=-\tr_{2}\left(\rho_{2}\log_{2}\rho_{2}\right),
\end{equation}
where $\rho_{1}=\tr_{2}\rho$ and $\rho_{2}=\tr_{1}\rho$ are the
reduced density matrices, $\rho=|\Psi\rangle\langle\Psi|$ is the
density matrix itself and $\tr_{i}$ denotes the partial trace over
the $i^{\textrm{th}}$ subsystem \cite{nielsen2011quantum}. Here
$E(\Psi)\in\left[0,1\right]$ and $E(\Psi)=0$ if and only if $|\Psi\rangle\in\mathcal{H}$
is not entangled.

\subsection{General entanglement content\label{sub:DefinitionOfH} }

Consider the Hamiltonian
\begin{equation}
H=\left(\begin{array}{cc}
re^{\ii\Theta} & s\\
s & re^{-\ii\Theta}
\end{array}\right),\quad r,s,\Theta\in\mathbb{R},\label{eq:DefinitionSimpleHamiltonian}
\end{equation}
with $s^{2}\geqslant r^{2}\sin^{2}\Theta$. Observe that $H$ is in
general not represented by a Hermitian matrix, $H^{\dagger}\neq H$.
But it obeys $\mathcal{PT}$-symmetry with
\begin{equation}
\mathcal{P}=\left(\begin{array}{cc}
0 & 1\\
1 & 0
\end{array}\right)
\end{equation}
and $\mathcal{T}$ complex conjugation \cite{PhysRevLett.89.270401}
(see also erratum \cite{PhysRevLett.92.119902}), i.e.~$\left[H,\mathcal{PT}\right]=0$.

Define $\sin\varphi\equiv r/s\cdot\sin\Theta$ with $\varphi\in\left[-\pi/2,\pi/2\right]$.
The simultaneous eigenstates of $H$ and $\mathcal{PT}$ are given
by
\begin{align}
|\phi_{+}\rangle & =\frac{1}{\sqrt{2\cos\varphi}}\left(\begin{array}{c}
e^{\ii\varphi/2}\\
e^{-\ii\varphi/2}
\end{array}\right),\nonumber \\
|\phi_{-}\rangle & =\frac{\ii}{\sqrt{2\cos\varphi}}\left(\begin{array}{c}
e^{-\ii\varphi/2}\\
-e^{\ii\varphi/2}
\end{array}\right),
\end{align}
 with eigenvalues $E_{\pm}=r\cos\Theta\pm s\cos\varphi$. These eigenstates
are orthonormal with respect to the positive $\mathcal{CPT}$-inner
product $\langle\Psi|\Phi\rangle_{\mathcal{CPT}}\equiv\langle\Psi|_{\mathcal{CPT}}\cdot|\Phi\rangle$,
where $\langle\Psi|_{\mathcal{CPT}}\equiv\left(\mathcal{CPT}|\Psi\rangle\right)^{\intercal}$,
see \cite{PhysRevLett.89.270401}. For this Hamiltonian, the $\mathcal{C}$
operator has the form
\begin{equation}
\mathcal{C}=\left(\begin{array}{cc}
\ii\tan\varphi & \sec\varphi\\
\sec\varphi & -\ii\tan\varphi
\end{array}\right)\label{eq:C-Operator}
\end{equation}
with $\left[H,\mathcal{C}\right]=0$.

In the following we will consider the composite system $\mathcal{H}=\mathcal{H}_{1}\times\mathcal{H}_{2}$
of Hilbert spaces $\mathcal{H}_{i=1,2}$ with dynamics governed by
the Hamiltonian $H=H_{1}\otimes H_{2}$, where $H_{i}$ is of the
form of \eqref{eq:DefinitionSimpleHamiltonian} with $r_{i},s_{i},\Theta_{i}\in\mathbb{R}$,
$s_{i}^{2}\geqslant r_{i}^{2}\sin^{2}\Theta_{i}$, $\sin\varphi_{i}\equiv r_{i}/s_{i}\cdot\sin\Theta_{i}$
and $|\phi_{\pm}\rangle_{i}\in\mathcal{H}_{i}$ being the eigenstates
of $H_{i}$ for $i=1,2$. Our respective $\mathcal{C}$ operator reads
$\mathcal{C}_{1}\otimes\mathcal{C}_{2}$, where $\mathcal{C}_{i}$
is of the form of \eqref{eq:C-Operator} with corresponding parameters.
As
\begin{align}
\left[H_{1}\otimes H_{2},\mathcal{PT}\otimes\mathcal{PT}\right] & =\left[H_{1}\otimes H_{2},\mathcal{C}_{1}\otimes\mathcal{C}_{2}\right]\nonumber \\
 & =\left[\mathcal{C}_{1}\otimes\mathcal{C}_{2},\mathcal{PT}\otimes\mathcal{PT}\right]\nonumber \\
 & =0
\end{align}
holds, $H$ obeys $\mathcal{PT}$ symmetry and we can define a $\mathcal{CPT}$-inner
product as above. The decomposition of the operators means the same
direct-product form of the metric operator $\eta$ in \cite{1751-8121-42-5-055303}.
According to proposition~2 and proposition~3 in the same paper,
this permits a proper quantum mechanical description of the bipartite
system with a unitary time evolution.

Now consider the density matrix $\rho=|\Psi(t)\rangle\langle\Psi(t)|_{\mathcal{CPT}}$
of a system in the state
\begin{multline}
|\Psi(t)\rangle=\alpha(t)|\phi_{+}\rangle_{1}\otimes|\phi_{+}\rangle_{2}+\beta(t)|\phi_{+}\rangle_{1}\otimes|\phi_{-}\rangle_{2}\\
+\,\gamma(t)|\phi_{-}\rangle_{1}\otimes|\phi_{+}\rangle_{2}+\delta(t)|\phi_{-}\rangle_{1}\otimes|\phi_{-}\rangle_{2}
\end{multline}
with $|\Psi(t)\rangle\in\mathcal{H}$. The eigenvalues of $\rho_{1}(t)=\tr_{2}\rho(t)$
are given by $\lambda_{\pm}(t)=\frac{1}{2}\pm\frac{1}{2}\sqrt{1-\Xi(t)}$
with $\Xi(t)=4|\alpha(t)\delta(t)-\beta(t)\gamma(t)|^{2}$, which
is a simplification of the result in \cite{springerlink:10.1007/s12043-009-0101-0}.
Hence the entanglement content is
\begin{multline}
E(t)\equiv E(\Psi(t))\\
=-\lambda_{+}(t)\log_{2}\lambda_{+}(t)-\lambda_{-}(t)\log_{2}\lambda_{-}(t).
\end{multline}
 Note that $|\Psi(t)\rangle$ only separates if $\alpha(t_{0})\delta(t_{0})=\beta(t_{0})\gamma(t_{0})$
for $t=t_{0}$.

\subsection{The Einstein-Podolsky-Rosen state}

Consider the \textit{Einstein-Podolsky-Rosen} state from conventional
quantum mechanics and normalize them with respect to the $\mathcal{CPT}$-inner
product:
\begin{equation}
\mid\uparrow\rangle_{i}=\sqrt{\cos\varphi_{i}}\left(\begin{array}{c}
1\\
0
\end{array}\right),\quad\mid\downarrow\rangle_{i}=\sqrt{\cos\varphi_{i}}\left(\begin{array}{c}
0\\
1
\end{array}\right).
\end{equation}
 Overall normalization yields
\begin{equation}
|\Psi^{-}\rangle=\kappa\left(\mid\uparrow\rangle_{1}\otimes\mid\downarrow\rangle_{2}-\mid\downarrow\rangle_{1}\otimes\mid\uparrow\rangle_{2}\right)\label{eq:EPRState}
\end{equation}
 with $\kappa=\left[2\left(1-\sin\varphi_{1}\sin\varphi_{2}\right)\right]^{-1/2}$.
The eigenvalues of the reduced density matrix of the first subsystem
are given by
\begin{equation}
\lambda_{\pm}=\frac{1}{2}\pm\frac{\sin\varphi_{1}-\sin\varphi_{2}}{2\left(1-\sin\varphi_{1}\sin\varphi_{2}\right)}.
\end{equation}
Our results are in disagreement with \cite{springerlink:10.1007/s12043-009-0101-0}
mainly due to the author's use of non-$\mathcal{CPT}$-normalized
states.

\begin{figure}
\begin{centering}
\includegraphics[width=1\columnwidth]{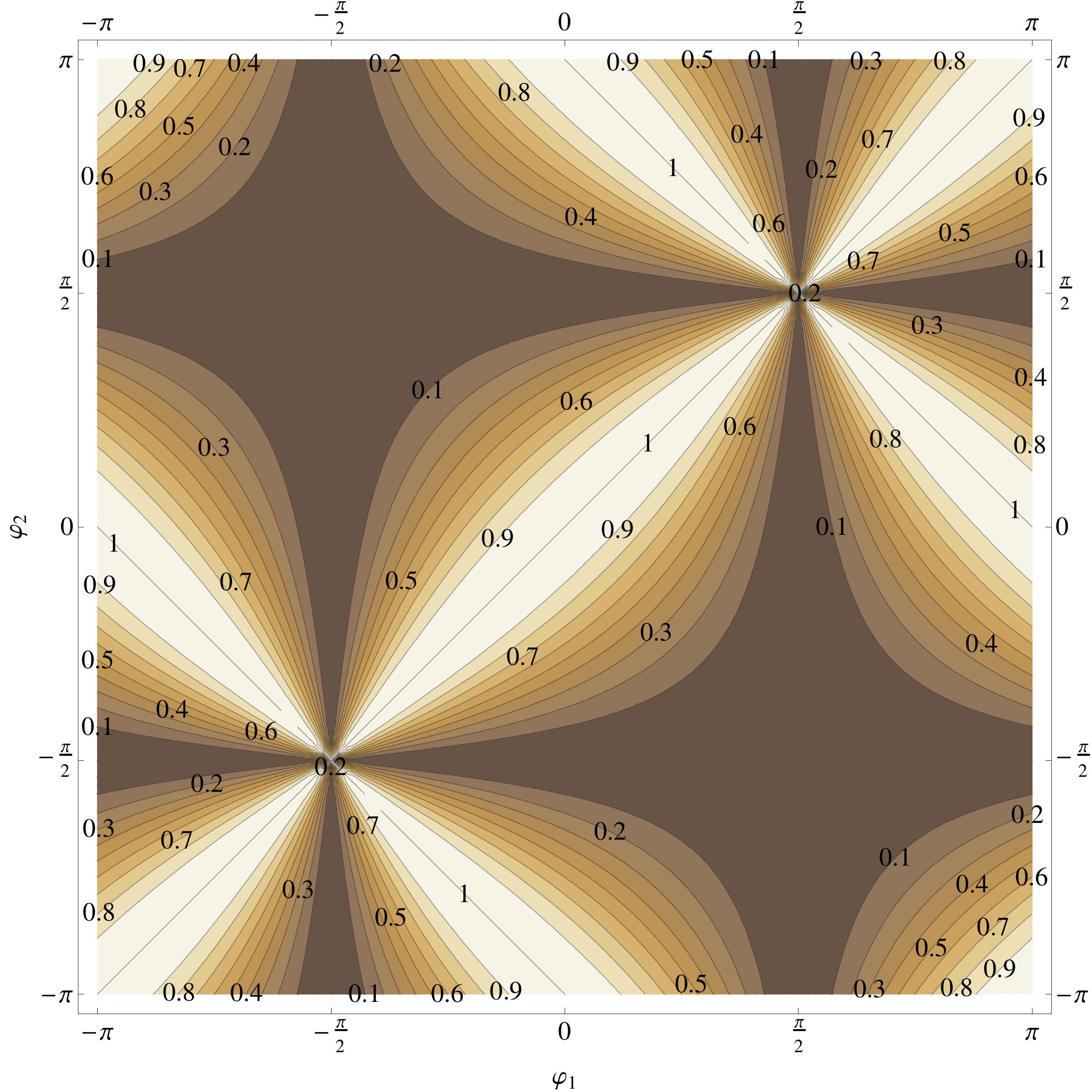} 
\par\end{centering}

\centering{}\caption{\textit{\label{fig:EntanglementContentPhi1Phi2}Entanglement content
$E(\Psi^{-})$ as a function of $\varphi_{1}$ and $\varphi_{2}$.}}
\end{figure}

We can consider the entanglement content
\begin{equation}
E(\Psi^{-})=-\lambda_{+}\log_{2}\lambda_{+}-\lambda_{-}\log_{2}\lambda_{-}
\end{equation}
as a function of $\varphi_{1}$ and $\varphi_{2}$, i.e.~of the Hamiltonians
$H_{1}$ and $H_{2}$. The result can be seen in figure \ref{fig:EntanglementContentPhi1Phi2}.
Note the cases $\varphi_{1}=\varphi_{2}$ or $\varphi_{1}+\varphi_{2}=\pm\pi$,
where \eqref{eq:EPRState} has the form of a $\mathcal{PT}$-symmetric
\textit{Bell} state.

\section{Entanglement generation}

\subsection{Entanglement capability}

We investigate the question how to increase the capability of a system
to entangle states. More precisely we want to understand the dependencies
of the entanglement capability $\Gamma(t)\equiv\textrm{d}E(t)/\textrm{d}t$
of the parameters of the system. In order to determine the time evolution
operator of the Hamiltonian \foreignlanguage{american}{$H_{1}\otimes H_{2}$}
(see section \ref{sub:DefinitionOfH}) define
\begin{equation}
\overrightarrow{n_{i}}\equiv\frac{2}{\omega_{i}}\left(s_{i},\,0,\,\ii r_{i}\sin\Theta_{i}\right)^{\intercal},\quad\omega_{i}\equiv2s_{i}\cos\varphi_{i}.\label{eq:DefinitionOfw1w2}
\end{equation}
Rewrite $H$ as
\begin{multline}
H=\left(r_{1}\cos\Theta_{1}\matrixOne+\frac{\omega_{1}}{2}\overrightarrow{n_{1}}\cdot\overrightarrow{\sigma}\right)\\
\otimes\left(r_{2}\cos\Theta_{2}\matrixOne+\frac{\omega_{2}}{2}\overrightarrow{n_{2}}\cdot\overrightarrow{\sigma}\right),
\end{multline}
with $\overrightarrow{\sigma}=\left(\sigma_{1},\sigma_{2},\sigma_{3}\right)^{\intercal}$,
$\sigma_{i}$ denoting the Pauli matrices and $\matrixOne$ the $2\times2$
identity matrix. Expanding the expression for $H$ yields four terms
of which only one can generate entanglement. We restrict ourselves
to this term and define $\tilde{H}=\omega_{1}\omega_{2}/4\cdot\left(\overrightarrow{n_{1}}\cdot\overrightarrow{\sigma}\otimes\overrightarrow{n_{2}}\cdot\overrightarrow{\sigma}\right)$
resulting in a time evolution operator $U(t)=\exp(-\ii\tilde{H}t/\hbar)$
given by
\begin{multline}
U(t)=\cos\left(\frac{\omega_{1}\omega_{2}t}{4\hbar}\right)\left(\matrixOne\otimes\matrixOne\right)\\
-\ii\sin\left(\frac{\omega_{1}\omega_{2}t}{4\hbar}\right)\left(\overrightarrow{n_{1}}\cdot\overrightarrow{\sigma}\otimes\overrightarrow{n_{2}}\cdot\overrightarrow{\sigma}\right).
\end{multline}
 
\begin{figure}
\begin{centering}
\includegraphics[width=1\columnwidth]{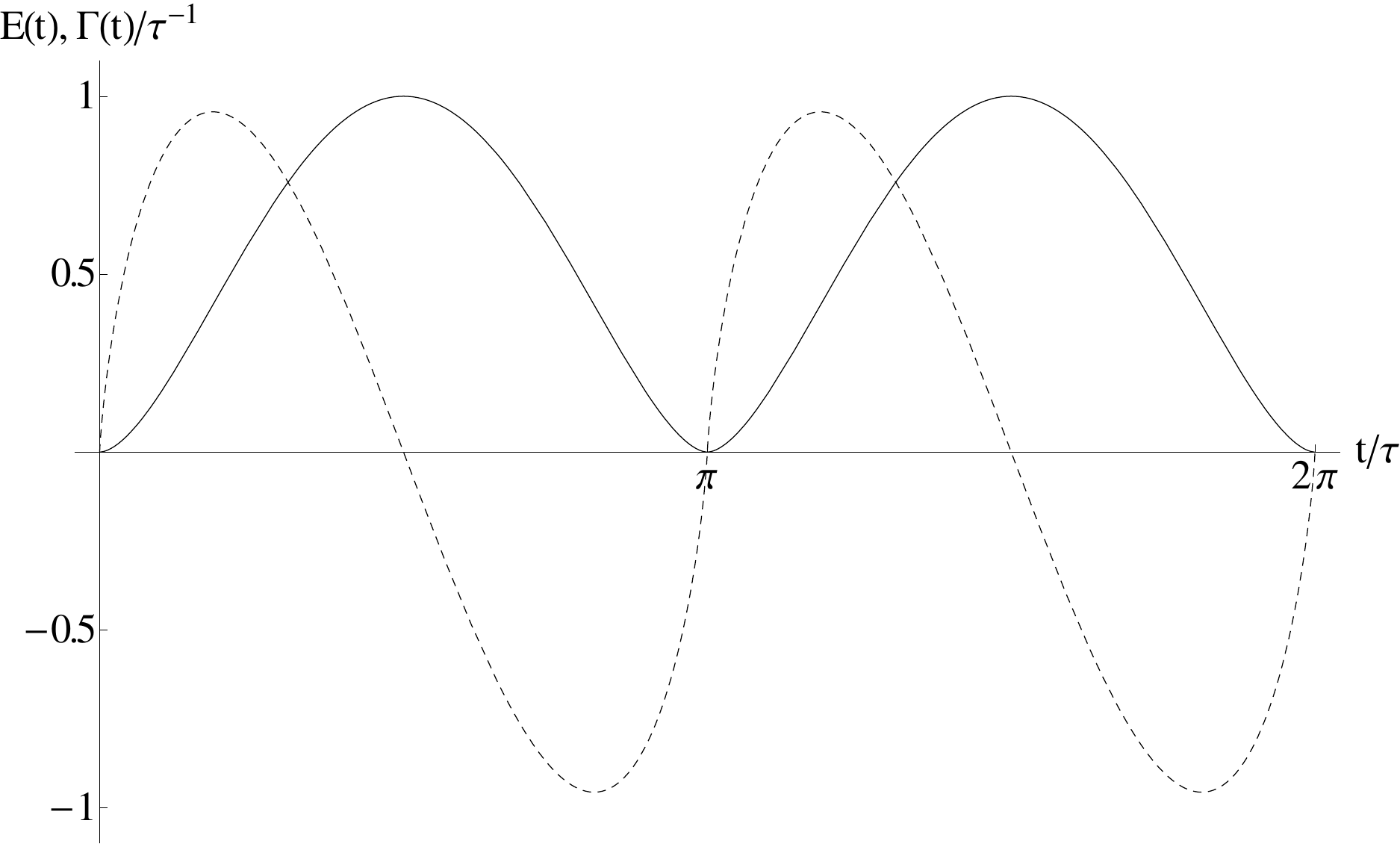} 
\par\end{centering}

\centering{}\caption{\label{fig:EntanglementContentAndRate}\textit{Entanglement content
$E(t)$ (solid) and entanglement rate $\Gamma(t)/\tau^{-1}$ (dashed)
of $|\Psi(t)\rangle$ with $\tau=2\hbar/\left(\omega_{1}\omega_{2}\right)$.}}
\end{figure}

Consider now a simple initial state, namely $|\Psi(t=0)\rangle=\,\mid\uparrow\rangle\otimes\mid\uparrow\rangle$,
and apply $U(t)$ to find
\begin{multline}
|\Psi(t)\rangle=\alpha(t)\mid\uparrow\rangle\otimes\mid\uparrow\rangle+\beta(t)\mid\uparrow\rangle\otimes\mid\downarrow\rangle\\
+\gamma(t)\mid\downarrow\rangle\otimes\mid\uparrow\rangle+\delta(t)\mid\downarrow\rangle\otimes\mid\downarrow\rangle
\end{multline}
with
\begin{align}
\alpha(t) & =\cos\left(\frac{\omega_{1}\omega_{2}t}{4\hbar}\right)+\frac{4\ii r_{1}r_{2}\sin\Theta_{1}\sin\Theta_{2}}{\omega_{1}\omega_{2}}\sin\left(\frac{\omega_{1}\omega_{2}t}{4\hbar}\right),\nonumber \\
\beta(t) & =\frac{4s_{2}r_{1}\sin\Theta_{1}}{\omega_{1}\omega_{2}}\sin\left(\frac{\omega_{1}\omega_{2}t}{4\hbar}\right),\nonumber \\
\gamma(t) & =\frac{4s_{1}r_{2}\sin\Theta_{2}}{\omega_{1}\omega_{2}}\sin\left(\frac{\omega_{1}\omega_{2}t}{4\hbar}\right),\nonumber \\
\delta(t) & =-\frac{4\ii s_{1}s_{2}}{\omega_{1}\omega_{2}}\sin\left(\frac{\omega_{1}\omega_{2}t}{4\hbar}\right).
\end{align}
Up to here our results are in agreement with \cite{springerlink:10.1007/s12043-009-0101-0}.
We now find the entanglement content to be given by
\begin{equation}
E(t)=-\lambda_{+}(t)\log_{2}\lambda_{+}(t)-\lambda_{-}(t)\log_{2}\lambda_{-}(t)
\end{equation}
with
\begin{equation}
\lambda_{\pm}(t)=\frac{1}{2}\pm\frac{1}{2}\cos\left(\frac{\omega_{1}\omega_{2}t}{2\hbar}\right).
\end{equation}
We find the entanglement rate to be
\begin{align}
\Gamma(t) & =\frac{\textrm{d}E(t)}{\textrm{d}t}\nonumber \\
 & =\frac{\omega_{1}\omega_{2}}{\hbar}\sin\left(\frac{\omega_{1}\omega_{2}t}{2\hbar}\right)\log_{16}\cot^{2}\left(\frac{\omega_{1}\omega_{2}t}{4\hbar}\right).
\end{align}
The maximal entanglement capability is $\Gamma_{\textrm{max}}=\max_{t}\Gamma(t)=0.4781\,\omega_{1}\omega_{2}/\hbar$.
Therefore by changing $\omega_{i}$ due to an adjustment of the parameters
of the Hamiltonians $H_{i}$, we can control \foreignlanguage{american}{$\Gamma_{\textrm{max}}$}.
The typical time dependency of the entanglement content and entanglement
rate can be seen in figure \ref{fig:EntanglementContentAndRate}.

\subsection{Efficiency of general systems}

If we have a given Hamiltonian one may ask how to maximize the entanglement
rate of the system. We want to generalize some results of conventional
quantum mechanics addressed in \cite{PhysRevLett.87.137901} to the
$\mathcal{PT}$-symmetric case.

Consider an arbitrary $\mathcal{PT}$-symmetric two qubit system.
Using the \textit{Schmidt}-de\-com\-po\-si\-tion theorem we rewrite
an arbitrary state $|\Psi(t)\rangle\in\mathcal{H}_{1}\times\mathcal{H}_{2}$
as
\begin{multline}
|\Psi(t)\rangle=\sqrt{p(t)}|\varphi_{t}\rangle\otimes|\chi_{t}\rangle\\
+e^{\ii\alpha}\sqrt{1-p(t)}|\varphi_{t}^{\perp}\rangle\otimes|\chi_{t}^{\perp}\rangle
\end{multline}
with $p\in\left[0,1\right]$ as \textit{Schmidt}-coefficient and $\langle\varphi_{t}|\varphi_{t}^{\perp}\rangle_{\mathcal{CPT}}=\langle\chi_{t}|\chi_{t}^{\perp}\rangle_{\mathcal{CPT}}=0$
as \textit{Schmidt}-vectors. The entanglement content is given by
\begin{equation}
E(t)=-p(t)\log_{2}p(t)-\left(1-p(t)\right)\log_{2}\left(1-p(t)\right)
\end{equation}
and the entanglement rate factorizes in two terms, i.e.~$\Gamma(t)=\textrm{d}E(t)/\textrm{d}p(t)\times\textrm{d}p(t)/\textrm{d}t$,
where
\begin{equation}
\frac{\textrm{d}E(t)}{\textrm{d}p(t)}=\frac{2}{\log2}\textrm{arctanh}\left(1-2p(t)\right).
\end{equation}
After choosing the phase $\alpha$ appropriately the evolution of
the \textit{Schmidt}-coefficient is determined by the differential
equation
\begin{multline}
\frac{\textrm{d}p(t)}{\textrm{d}t}=\frac{2}{\hbar}\sqrt{p(t)\left(1-p(t)\right)}\\
\times|\left(\langle\varphi_{t}|_{\mathcal{CPT}}\otimes\langle\chi_{t}|_{\mathcal{CPT}}\right)H\left(|\varphi_{t}^{\perp}\rangle\otimes|\chi_{t}^{\perp}\rangle\right)|.
\end{multline}
This relation is known from conventional quantum mechanic, but also
holds for $\mathcal{PT}$-symmetric systems. Define
\begin{equation}
\hbar\Omega\equiv\max_{\substack{\|\varphi\|=1\\
\|\chi\|=1
}
}|\left(\langle\varphi|_{\mathcal{CPT}}\otimes\langle\chi|_{\mathcal{CPT}}\right)H\left(|\varphi^{\perp}\rangle\otimes|\chi^{\perp}\rangle\right)|.\label{eq:DefinitionOfOmega}
\end{equation}
Then the time evolution of $p(t)$ for an optimally prepared setup,
i.e.~a setup forcing the qubit states to be optimal at every instant
of time
\begin{equation}
|\left(\langle\varphi_{t}|_{\mathcal{CPT}}\otimes\langle\chi_{t}|_{\mathcal{CPT}}\right)H\left(|\varphi_{t}^{\perp}\rangle\otimes|\chi_{t}^{\perp}\rangle\right)|=\hbar\Omega
\end{equation}
for all $t$, follows from the differential equation
\begin{equation}
\frac{\textrm{d}p_{\textrm{opt}}(t)}{\textrm{d}t}=2\Omega\sqrt{p_{\textrm{opt}}(t)\left(1-p_{\textrm{opt}}(t)\right)}.
\end{equation}
We find $p_{\textrm{opt}}(t)=\sin^{2}\left(\Omega t+\delta_{0}\right)$
with an integration constant $\delta_{0}$, which is in agreement
with results from conventional quantum mechanics in \cite{PhysRevLett.87.137901}
(see also \cite{springerlink:10.1007/s12043-009-0101-0}). Hence we
can characterize the entanglement rate of an optimally prepared setup
completely in terms of $\Omega$ via
\begin{align}
\Gamma_{\textrm{opt}}(t) & =\frac{\textrm{d}p_{\textrm{opt}}(t)}{\textrm{d}t}\cdot\frac{\textrm{d}E(t)}{\textrm{d}p(t)}\Biggl|_{p_{\textrm{opt}}(t)}\\
 & =\frac{2\Omega}{\log2}\textrm{arctanh}\left(\cos\left(2\Omega t+2\delta_{0}\right)\right)\sin\left(2\Omega t+2\delta_{0}\right).\nonumber 
\end{align}
and find for the maximal rate $\Gamma_{\textrm{max}}=\max_{t}\Gamma_{\textrm{opt}}(t)=1.9123\,\Omega$.
The parameter $\Omega$ is the critical value one needs to maximize
to efficiently entangle states. We remark that this result also holds
in conventional quantum mechanics, where $\Omega$ is defined with
conventional conjugates instead of $\mathcal{CPT}$-conjugates.

\section{Conclusions}

For the case of a state from a bipartite system we determined the
degree of entanglement and saw the emergence of symmetrical patterns
(see figure \ref{fig:EntanglementContentPhi1Phi2}) in the case of
the \textit{Einstein-Podolsky-Rosen} state. We quantified the capability
of a given \textit{$\mathcal{PT}$}-symmetric system to generate entangled
states in terms of the parameters of the Hamiltonian. Their ability
to entangle states can be described by the parameters $\omega_{i=1,2}$
in \eqref{eq:DefinitionOfw1w2} or in general by $\Omega$ in \eqref{eq:DefinitionOfOmega}.
Many relations are similar to the corresponding ones in conventional
quantum mechanics after replacing the usual inner product with the
\textit{$\mathcal{CPT}$}-inner product. However, these results are
not obvious and need to be checked. For example, the recent discussion
of the quantum brachistochrone problem showed that \textit{$\mathcal{PT}$}-symmetric
quantum mechanics can give some surprising results \cite{PhysRevLett.98.040403,PhysRevA.78.042115,PhysRevLett.101.230404,PhysRevA.79.014101}.\medskip{}

\begin{acknowledgments}
This work was carried out at the Department of Physics of the National
University of Singapore, whose hospitality is gratefully acknowledged.
The results were obtained within the UROPS program and we thank the
National University of Singapore, the University of Heidelberg and
the German Academic Foundation for financial support.
\end{acknowledgments}
\vfill{}

\noindent \bibliographystyle{siam-url}
\bibliography{PTLiterature}

\end{document}